The following is the entire tex file
which should be compiled with the standard latex macro.
There is no figure.
----------------------------------------------------------------------
\documentstyle[12pt]{article}
\newcommand{\bra}{\langle}
\newcommand{\ket}{\rangle}
\newcommand{\k}{\mbox{i}}

\makeatletter
\@addtoreset{equation}{section}
\makeatother

\begin{document} 
%
%
\begin{flushright}
DPNU-92-39 \\
September 1992
\end{flushright}
\begin{center}
\vspace*{36mm}
{\Large \bf Relativistic Generalization and}
\\
\vspace{10pt}
{\Large \bf Extension to the Non-Abelian Gauge Theory}
\\
\vspace{10pt}
{\Large \bf of Feynman's Proof of the Maxwell Equations}
\footnote[2]{This article has been published in {\it Annals of Physics}
{\bf 220}(1992), 229-247.}
\\
\vspace{16pt}
{\large Shogo Tanimura}
\footnote[3]{e-mail address : tanimura@eken.phys.nagoya-u.ac.jp}
\\
\vspace{10pt}
{\large \it Department of Physics, Nagoya University, }
\\
{\large \it Nagoya 464-01, Japan}
\\
%
%
\vspace{24pt}
{\bf Abstract}
\\
\vspace{12pt}
\begin{minipage}[t]{130mm}
\hspace{10pt}
\baselineskip 6mm
 R.P. Feynman showed F.J. Dyson a proof of the Lorentz force law
and the homogeneous Maxwell equations,
which he obtained starting from Newton's law of motion
and the commutation relations between position and velocity
for a single nonrelativistic particle.
We formulate both a special relativistic and a general relativistic versions
of Feynman's derivation.
Especially in the general relativistic version
we prove that the only possible fields
that can consistently act on a quantum mechanical particle
are scalar, gauge and  gravitational fields.
We also extend Feynman's scheme to the case of non-Abelian gauge theory
in the special relativistic context.
\end{minipage}
\end{center}
%
%
\newpage
\baselineskip 7mm
\section{Introduction}
\hspace{12pt}
In 1990 F.J. Dyson published a paper \cite{Dyson2}
about one of R.P. Feynman's works,
the proof of the Lorentz force law and the homogeneous Maxwell equations.
According to Dyson \cite{Dyson1}, Feynman showed Dyson the proof in 1948.
Feynman started with the commutation relations
between position and velocity of a single nonrelativistic particle
obeying Newton's law of motion
and deduced the existence of electric and magnetic fields
satisfying the equations of Maxwell.
However he had never published his proof.
After Feynman's death,
Dyson published it with some editorial comments.
Thanks to Dyson, Feynman's work is now available to us.
\par
Feynman's proof is mathematically rigorous.
However relativistic covariance is not manifest in his proof.
In this paper we propose both a special and a general
relativistic generalization of it.
Especially in the general relativistic version we show
that the only possible fields that can consistently act
on a quantum mechanical particle are scalar, gauge and gravitational fields.
We also extend Feynman's scheme to the case of non-Abelian gauge theory.
We add some remarks to each case.
%
%
\section{Review of Feynman's proof}
\hspace{12pt}
  First we review Feynman's proof of the Maxwell equations following
  Dyson \cite{Dyson2}.
\\
Their assumptions :
\begin{enumerate}
	\renewcommand{\labelenumi}{(\roman{enumi})}
	\item
		A particle is moving in a 3-dimensional Euclidean space
		with position $x_i(t)$ $(i=1,2,3)$, where $t$ is time.
	\item
		Its position and velocity $\dot x_i(t)$
		satisfy the commutation relations
		\begin{eqnarray}
			&&
			[ \: x_i, x_j \: ] = 0,
		\label{eqn:2.1}
			\\ &&
			m \, [ \: x_i, \dot x_j \: ]
			= \mbox{i} \hbar \, \delta_{ij}.
		\label{eqn:2.2}
		\end{eqnarray}
	\item
		It obeys the equation of motion
		\begin{equation}
			m \: \ddot x_i = F_i (x,\dot x,t).
		\label{eqn:2.3}
		\end{equation}
\end{enumerate}
Their results :
\begin{enumerate}
	\renewcommand{\labelenumi}{(\roman{enumi})}
	\item
		The force $F_i(x,\dot x,t)$ can be written as
		\begin{equation}
			F_i(x,\dot x,t) =
			E_i(x,t)
			+ \epsilon_{ijk} \bra \dot x_j B_k(x,t) \ket,
		\label{eqn:2.4}
		\end{equation}
		where the symbol $ \bra \cdots \ket $ refers to
		the Weyl-ordering
		prescription.
		Later we shall give explanation on this prescription.
	\item
		The fields $E_i(x,t)$ and $B_i(x,t)$ satisfy
		the homogeneous Maxwell equations
		\begin{eqnarray}
			&&
			\mbox{div} \: B = 0 ,
		\label{eqn:2.5}
			\\ &&
			\frac{\partial B}{\partial t}
			+ \mbox{rot} \: E = 0,
		\label{eqn:2.6}
		\end{eqnarray}
		which implies that
		there exist a scalar potential $\phi(x,t)$
		and a vector potential $ A_i(x,t) $ such that
		\begin{eqnarray}
			&&
			B = \mbox{rot} \: A,
		\label{eqn:2.7}
			\\ &&
			E = - \,\mbox{grad}\:\phi - \mbox{rot}\: A.
		\label{eqn:2.8}
		\end{eqnarray}
\end{enumerate}
{}From these results
they identified $E$ and $B$ with electric and magnetic field respectively.
Dyson appreciated the point that the proof shows that
the only possible fields that can consistently act
on a quantum mechanical particle are gauge fields.
\par
Their proof : Differentiating equation (\ref{eqn:2.2})
with respect to time and using (\ref{eqn:2.3}), one obtains
\begin{equation}
	m \, [ \: \dot x_i, \dot x_j \: ] + [ \: x_i , F_j \: ] = 0 .
\label{eqn:2.9}
\end{equation}
This allows one to write
\begin{equation}
	m \, [ \: \dot x_i, \dot x_j \: ]
	= - [ \: x_i , F_j \: ]
	= \frac{\k\hbar}{m} \: \epsilon_{ijk} \: B_k .
\label{eqn:2.10}
\end{equation}
One may consider this equation as the definition of the field $B$.
Equation (\ref{eqn:2.10}) may be written as
\begin{equation}
	B_k =
	\frac{m^2}{2\,\k\hbar} \: \epsilon_{klm} \:
	[ \: \dot x_l, \dot x_m \: ].
\label{eqn:2.11}
\end{equation}
The field $B$ would depend on $x , \dot x $ and $t$.
But the Jacobi identity and Eq. (\ref{eqn:2.2}) imply
\begin{eqnarray}
	[ \: x_i , B_k \: ]
	&=&
	\frac{m^2}{2\,\k\hbar} \: \epsilon_{klm} \:
	[\: x_i,[\:\dot x_l,\dot x_m \:]\:]
	\nonumber\\
	&=&
	\frac{m^2}{2\,\k\hbar} \: \epsilon_{klm} \:
	\Bigl(
		[\: [\: x_i, \dot x_l \:], \dot x_m \:] +
		[\: \dot x_l, [\: x_i , \dot x_m \:]\:]
	\Bigr)
	\nonumber\\
	&=&
	0,
\label{eqn:2.12}
\end{eqnarray}
which means that $B$ is a function of $x$ and $t$ only.
For a function $ f(x,t) $ one has the formula
\begin{equation}
	[\: \dot x_k , f(x,t) \:] =
	- \frac{\k\hbar}{m} \: \frac{\partial f}{\partial x_k},
\label{eqn:lemma}
\end{equation}
which is easily verified by Eqs. (\ref{eqn:2.1}), (\ref{eqn:2.2}).
One can use Eq. (\ref{eqn:lemma}) to obtain
\begin{equation}
	[\: \dot x_k , B_k \:]
	= - \frac{\k\hbar}{m} \: \frac{\partial B_k}{\partial x_k}.
\label{eqn:2.13}
\end{equation}
On the other hand Eq. (\ref{eqn:2.11}) and the Jacobi identity give
\begin{equation}
	[\: \dot x_k , B_k \:]
	=
	\frac{m^2}{2\,\k\hbar} \: \epsilon_{klm} \:
	[\: \dot x_k ,[ \: \dot x_l, \dot x_m \: ] \:]
	=
	0.
\label{eqn:2.14}
\end{equation}
These prove (\ref{eqn:2.5}).
\par
Next one takes Eq. (\ref{eqn:2.4}) as the definition of the field $E$.
Here it is necessary to explain the Weyl-ordering.
It refers to complete symmetrization of operator-products, for instance,
\begin{eqnarray*}
	&&
	\bra x_i \dot x_j \ket
	= \frac12 \, ( x_i \dot x_j + \dot x_j x_i ),
	\\
	&&
	\bra x_i \dot x_j \dot x_k \ket
	= \frac16 \, ( x_i \dot x_j \dot x_k
	             + x_i \dot x_k \dot x_j
	             + \dot x_j x_i \dot x_k
	             + \dot x_j \dot x_k x_i
	             + \dot x_k x_i \dot x_j
	             + \dot x_k \dot x_j x_i  )
\end{eqnarray*}
and so on.
Again, $E$ would depend on $x, \dot x$ and $t$ in general,
but using Eqs. (\ref{eqn:2.4}), (\ref{eqn:2.10}),
(\ref{eqn:2.2}) and (\ref{eqn:2.12}) in this order, one obtains
\begin{eqnarray}
	[\: x_l , E_i \:]
	&=&
	[\: x_l , F_i \:] - \epsilon_{ijk} \:
	\bra [\: x_l,\dot x_j \:] \: B_k + \dot x_j \: [\: x_l,B_k \:] \ket
	\nonumber\\
	&=&
	- \frac{\k\hbar}{m} \: \epsilon_{lik} \: B_k
	- \epsilon_{ijk} \: \frac{\k\hbar}{m} \: \delta_{lj} \, B_k
	\nonumber\\
	&=& 0 ,
\label{eqn:2.15}
\end{eqnarray}
which says that $E$ is a function of $x$ and $t$ only.
\par
It remains to prove the second Maxwell equation (\ref{eqn:2.6}).
One takes the total time-derivative of (\ref{eqn:2.11}) and obtains
\begin{equation}
	\frac{\partial B_k}{\partial t}
	+ \bra \dot x_j \frac{\partial B_k}{\partial x_j} \ket
	=
	\frac{m^2}{\k\hbar} \: \epsilon_{klm} \: [\: \dot x_l,\ddot x_m \:].
\label{eqn:2.16}
\end{equation}
Now by (\ref{eqn:2.3}), (\ref{eqn:2.4}), (\ref{eqn:2.10}),
(\ref{eqn:lemma}) and (\ref{eqn:2.5}),
the right side of (\ref{eqn:2.16}) becomes
\begin{eqnarray}
	&&
	\frac{m}{\k\hbar} \: \epsilon_{klm} \: [\: \dot x_l ,
	E_m + \epsilon_{mij} \: \bra \dot x_i B_j \ket \:]
	\nonumber\\
	=
	&&
	\frac{m}{\k\hbar} \:
	\Bigl(
	\epsilon_{klm} \:[\: \dot x_l , E_m \:]
	+ ( \delta_{ik}\,\delta_{jl} - \delta_{il}\,\delta_{jk} )
	\: \bra \,
	[\: \dot x_l,\dot x_i \:] \: B_j + \dot x_i \: [\: \dot x_l,B_j \:]
	\, \ket
	\Bigr)
	\nonumber\\
	=
	&&
	- \epsilon_{klm} \frac{\partial E_m}{\partial x_l}
	+ ( \delta_{ik}\,\delta_{jl} - \delta_{il}\,\delta_{jk} )
	\: \bra \,
	\frac 1m \epsilon_{lin} \: B_n \, B_j
	- \dot x_i \: \frac{\partial B_j}{\partial x_l}
	\, \ket
	\nonumber\\
	=
	&&
	- \epsilon_{klm} \frac{\partial E_m}{\partial x_l}
	+
	\bra \,
	\frac 1m \epsilon_{jkn} \: B_n \, B_j
	- \dot x_k \: \frac{\partial B_j}{\partial x_j}
	+ \dot x_i \: \frac{\partial B_k}{\partial x_i}
	\, \ket
	\nonumber\\
	=
	&&
	- \epsilon_{klm} \frac{\partial E_m}{\partial x_l}
	+ \bra \dot x_i \: \frac{\partial B_k}{\partial x_i} \ket .
\label{eqn:2.17}
\end{eqnarray}
Eqs. (\ref{eqn:2.16}) and (\ref{eqn:2.17}) give (\ref{eqn:2.6}).
End of proof.
\par
Remark : We observe the properties of the commutator
$ [ \; , \; ] $ used in their argument.
They are
\begin{enumerate}
	\renewcommand{\labelenumi}{(\roman{enumi})}
	\item
		bilinearity
		\begin{eqnarray}
		&&
		[\: \lambda A + \mu B,C \:]
		= \lambda \, [\: A,C \:]
		+ \mu     \, [\: B,C \:],
		\label{eqn:bilinearity}
		\\
		&&
		[\: A, \lambda B + \mu C \:]
		= \lambda \, [\: A,B \:]
		+ \mu     \, [\: A,C \:],
		\end{eqnarray}
	\item
		antisymmetry
		\begin{equation}
		[\: A,B \:] = - [\: B,A \:],
		\end{equation}
	\item
		Jacobi identity
		\begin{equation}
		[\: A,[\: B,C \:]\:]+
		[\: B,[\: C,A \:]\:]+
		[\: C,[\: A,B \:]\:]=0	,
		\end{equation}
	\item
		Leibniz rule I
		\begin{equation}
		[\: A,BC \:] = [\: A,B \:] \, C + B \, [\: A,C \:],
		\end{equation}
	\item
		Leibniz rule II
		\begin{equation}
		\frac{\mbox{d}}{\mbox{d}t}[\: A,B \:]
		= [\: \frac{\mbox{d}A}{\mbox{d}t},B \:]
		+ [\: A,\frac{\mbox{d}B}{\mbox{d}t} \:].
		\label{eqn:Leibniz2}
		\end{equation}
\end{enumerate}
The Poisson bracket $ \{ \; , \; \} $ also possesses the properties
from (i) to (iv) as automatic consequence of its definition.
However the property (v) is not trivial for the Poisson bracket.
Unless the canonical equation of motion is given,
the Poisson bracket does not satisfy the property (v).
We shall show what could be the matter.
\par
We treat a classical mechanical system of one degree of freedom for simplicity.
$ A(q,p) $ and $ B(q,p) $ are observables
which do not depend on time explicitly.
The Poisson bracket is defined by
\begin{equation}
	\{ A,B \} =
	\frac{\partial A}{\partial q} \: \frac{\partial B}{\partial p} -
	\frac{\partial A}{\partial p} \: \frac{\partial B}{\partial q}   .
\label{eqn:Poisson}
\end{equation}
Take the total time-derivative of Eq. (\ref{eqn:Poisson})
through time-dependence of $ q(t) $ and $ p(t) $. We obtain
\begin{equation}
	\frac{\mbox{d}}{\mbox{d}t} \{ A,B \}
	= \{ \frac{\mbox{d}A}{\mbox{d}t},B \}
	+ \{ A,\frac{\mbox{d}B}{\mbox{d}t} \}
	- \{ A,B \}
	\left(
	       \frac{\partial \dot q}{\partial q}  +
	       \frac{\partial \dot p}{\partial p}
	\right).
\label{eqn:Leibniz?}
\end{equation}
Without the equation of motion
$ \dot q = f(q,p), \: \dot p = g(q,p) $,
we can say nothing about the last term of Eq. (\ref{eqn:Leibniz?}).
If we have the Hamiltonian $ H(q,p) $ and use the canonical equation of motion
\begin{eqnarray}
	&&
	\dot q =   \frac{\partial H}{\partial p},
	\nonumber
	\\
	&&
	\\
	&&
	\dot p = - \frac{\partial H}{\partial q},
	\nonumber
\end{eqnarray}
the last term of Eq. (\ref{eqn:Leibniz?}) vanishes and
the Leibniz rule II is satisfied.
It is one of virtues of Feynman's proof that
there is no need of a priori existence of Hamiltonian,
Lagrangian, canonical equation or Heisenberg equation.
%
%
\section{Special relativistic version}
\hspace{12pt}
It is a weak point of Feynman's derivation that
Lorentz covariance is not manifest \cite{comments} \cite{Vaidya}.
We propose a special relativistic version of it.
\\
Assumptions :
\begin{enumerate}
	\renewcommand{\labelenumi}{(\roman{enumi})}
	\item
		A particle is moving in $d$-dimensional Minkowski
		space-time with coordinate $x^\mu(\tau)$
		$(\mu = 0, 1, \cdots, d-1)$, where $\tau$ is a parameter.
	\item
		Its coordinate and velocity $\dot x^\mu(\tau)$
		satisfy the commutation relations
		\begin{eqnarray}
			&&
			[ \: x^\mu, x^\nu \: ] = 0,
		\label{eqn:3.1}
			\\ &&
			m \, [ \: x^\mu, \dot x^\nu \: ]
			= - \mbox{i} \hbar \, \eta^{\mu \nu},
		\label{eqn:3.2}
		\end{eqnarray}
		where the dot refers to the derivative with respect to $\tau$,
		and $\eta^{\mu \nu}$ is Minkow\-ski\-an metric
		$ \eta = \mbox{diag} (+1,-1,\cdots,-1) $.
	\item
		It obeys the equation of motion
		\begin{equation}
			m \: \ddot x^\mu = F^\mu (x,\dot x).
		\label{eqn:3.3}
		\end{equation}
\end{enumerate}
Results :
\begin{enumerate}
	\renewcommand{\labelenumi}{(\roman{enumi})}
	\item
		The force $F^\mu(x,\dot x)$ can be written as
		\begin{equation}
			F^\mu(x,\dot x) =
			G^\mu(x) +
			\bra F^\mu_{\:\;\;\nu}(x) \, \dot x^\nu \ket,
		\label{eqn:3.4}
		\end{equation}
		where the symbol $ \bra \cdots \ket $ also refers to
		the Weyl-ordering prescription.
	\item
		The fields $G^\mu(x), F^\mu_{\;\;\nu}(x)$ satisfy
		\begin{eqnarray}
			&&
			\partial_\mu G_\nu - \partial_\nu G_\mu = 0,
		\label{eqn:3.5}
			\\ &&
			\partial_\mu F_{\nu \rho} +
			\partial_\nu F_{\rho \mu} +
			\partial_\rho F_{\mu \nu} =0,
		\label{eqn:3.6}
		\end{eqnarray}
		which implies that there exist a scalar field $\phi(x)$
		and a vector field $ A_\mu(x) $ such that
		\begin{eqnarray}
			&&
			G_\mu = \partial_\mu \phi,
		\label{eqn:3.7}
			\\ &&
			F_{\mu \nu}=\partial_\mu A_\nu-\partial_\nu A_\mu,
		\label{eqn:3.8}
		\end{eqnarray}
\end{enumerate}
\par
Proof :
Differentiating Eq. (\ref{eqn:3.2}) with respect to $\tau$
and using (\ref{eqn:3.3}), we obtain
\begin{equation}
	m \, [ \: \dot x^\mu, \dot x^\nu \: ] + [ \: x^\mu , F^\nu \: ] = 0 .
\label{eqn:3.9}
\end{equation}
We define $ F^{\mu \nu} $ by
\begin{equation}
	m \, [ \: \dot x^\mu, \dot x^\nu \: ]
	= - [ \: x^\mu , F^\nu \: ]
	= - \frac{\k\hbar}m F^{\mu \nu} .
\label{eqn:3.10}
\end{equation}
By definition, $ F^{\mu \nu} = - F^{\nu \mu} $.
$ F^{\mu \nu} $ would depend on $x$ and $\dot x$.
But using the Jacobi identity and Eq. (\ref{eqn:3.2}) we get
\begin{eqnarray}
	[ \: x^\lambda , F^{\mu \nu} \: ]
	&=&
	- \frac{m^2}{\k\hbar} \:
	[\: x^\lambda,[\:\dot x^\mu, \dot x^\nu \:]\:]
	\nonumber\\
	&=&
	- \frac{m^2}{\k\hbar} \:
	\Bigl(
		[\: [\: x^\lambda, \dot x^\mu \:], \dot x^\nu \:] +
		[\: \dot x^\mu, [\: x^\lambda , \dot x^\nu \:]\:]
	\Bigr)
	\nonumber\\
	&=&
	0,
\label{eqn:3.11}
\end{eqnarray}
which means that $F^{\mu \nu}$ is a function of $x$ only.
\par
Here we should pay attention to the fact that raising and lowering
of tensor-indices by $ \eta^{\mu \nu} $ and $ \eta_{\mu \nu} $
are compatible with operator-product.
For example if we define $ \dot x_\mu $ and $ F_{\nu \rho} $ as
\begin{eqnarray}
	&&
	\dot x_\mu = \eta_{\mu \alpha} \, \dot x^\alpha ,
\label{eqn:3.12}
	\\
	&&
	F_{\nu \rho} = \eta_{\nu \alpha}
	            \, \eta_{\rho \beta}
	            \, F^{\alpha \beta} ,
\label{eqn:3.13}
\end{eqnarray}
respectively, it is justified to write
\begin{equation}
	F_{\nu \rho} =
	- \frac{m^2}{\k\hbar} \:[\: \dot x_\nu, \dot x_\rho \:].
\label{eqn:3.14}
\end{equation}
Eqs. (\ref{eqn:3.2}), (\ref{eqn:3.12}) imply
\begin{equation}
	m \, [ \: x^\mu, \dot x_\nu \: ]
	= - \mbox{i} \hbar \, \delta^\mu_{\;\;\nu} ,
\label{eqn:3.15}
\end{equation}
from which we derive a useful formula
\begin{equation}
	[\: \dot x_\nu, f(x) \:] =
	\frac{\k\hbar}m \: \frac{\partial f}{\partial x^\nu}
\label{eqn:lemma2}
\end{equation}
for a function $ f(x) $.
\par
The Jacobi identity with (\ref{eqn:3.14}) and (\ref{eqn:lemma2}) implies
\begin{eqnarray}
	0
	&=&
	[\: \dot x_\mu , [\: \dot x_\nu , \dot x_\rho \:] \:] +
	[\: \dot x_\nu , [\: \dot x_\rho, \dot x_\mu  \:] \:] +
	[\: \dot x_\rho, [\: \dot x_\mu , \dot x_\nu  \:] \:]
	\nonumber\\
	&=&
	- \frac{\k\hbar}{m^2} \,
	\Bigl(
	[\: \dot x_\mu , F_{\nu  \rho} \:] +
	[\: \dot x_\nu , F_{\rho \mu } \:] +
	[\: \dot x_\rho, F_{\mu  \nu } \:]
	\Bigr)
	\nonumber\\
	&=&
	\frac{\hbar^2}{m^3} \,
	( \:
	\partial_\mu F_{\nu \rho} +
	\partial_\nu F_{\rho \mu} +
	\partial_\rho F_{\mu \nu}
	\: ),
\label{eqn:3.16}
\end{eqnarray}
which is nothing but Eq. (\ref{eqn:3.6}).
\par
Next we take Eq. (\ref{eqn:3.4}) as the definition of $ G^\mu $, that is,
\begin{equation}
	G^\mu = F^\mu (x,\dot x) - \bra F^{\mu \nu}(x) \, \dot x_\nu \ket.
\label{eqn:3.17}
\end{equation}
Again, $ G^\mu $ might depend on $x$ and $\dot x$,
but using Eqs. (\ref{eqn:3.10}), (\ref{eqn:3.11}) and (\ref{eqn:3.15}) we get
\begin{eqnarray}
	[\: x^\lambda, G^\mu \:]
	&=&
	[\: x^\lambda, F^\mu \:] -
	\bra F^{\mu \nu} [\: x^\lambda, \dot x_\nu \:] \ket
	\nonumber
	\\
	&=&
	\frac{\k\hbar}m \, F^{\lambda \mu} +
	\frac{\k\hbar}m \, F^{\mu \nu} \, \delta^\lambda_{\;\;\nu}
	\nonumber
	\\
	&=&
	0,
\label{eqn:3.18}
\end{eqnarray}
which says that $ G^\mu $ is also a function of $x$ only.
\par
It completes the proof to show equation (\ref{eqn:3.5}).
Eqs.(\ref{eqn:3.17}), (\ref{eqn:3.3}), (\ref{eqn:lemma2})
and (\ref{eqn:3.10}) imply
\begin{eqnarray*}
	&&
	[\: \dot x_\mu, G_\nu \:]
	\\
	= &&
	[\: \dot x_\mu, F_\nu \:] -
	\bra \: [\: \dot x_\mu, F_{\nu \rho} \:] \dot x^\rho \: \ket -
	\bra \: F_{\nu \rho} [\: \dot x_\mu, \dot x^{\rho} \:] \: \ket
	\\
	= &&
	m [\: \dot x_\mu, \ddot x_\nu \:] -
	\frac{\k\hbar}m
	\bra \: \partial_\mu F_{\nu \rho} \, \dot x^\rho \: \ket +
	\frac{\k\hbar}{m^2} F_{\nu \rho} \, F_\mu^{\;\:\rho},
\end{eqnarray*}
which leads to
\begin{eqnarray}
	&&
	[\: \dot x_\mu, G_\nu \:] -
	[\: \dot x_\nu, G_\mu \:]
	\nonumber
	\\
	= &&
	m [\: \dot x_\mu, \ddot x_\nu \:] -
	m [\: \dot x_\nu, \ddot x_\mu \:] -
	\frac{\k\hbar}m
	\bra \:
	( \partial_\mu F_{\nu \rho} - \partial_\nu F_{\mu \rho} ) \dot x^\rho
	\: \ket
	\nonumber
	\\
	&&
	+ \frac{\k\hbar}{m^2}
	( F_{\nu \rho} F_\mu^{\;\:\rho} - F_{\mu \rho} F_\nu^{\;\:\rho} )
	\nonumber
	\\
	= &&
	m \frac{\mbox{d}}{\mbox{d} \tau} [\: \dot x_\mu, \dot x_\nu \:] -
	\frac{\k\hbar}m
	\bra \:
	( \partial_\mu F_{\nu \rho} - \partial_\nu F_{\mu \rho} ) \dot x^\rho
	\: \ket
	\nonumber
	\\
	= &&
	- \frac{\k\hbar}m \frac{\mbox{d}}{\mbox{d} \tau} F_{\mu \nu}
	- \frac{\k\hbar}m
	\bra \:
	( \partial_\mu F_{\nu \rho} + \partial_\nu F_{\rho \mu} ) \dot x^\rho
	\: \ket
	\nonumber
	\\
	= &&
	- \frac{\k\hbar}m
	\bra \:
	( \partial_\rho F_{\mu \nu}
	+ \partial_\mu  F_{\nu \rho}
	+ \partial_\nu  F_{\rho \mu} ) \dot x^\rho
	\: \ket .
\label{eqn:3.19}
\end{eqnarray}
Eqs. (\ref{eqn:3.6}), (\ref{eqn:lemma2}) and (\ref{eqn:3.19})
give (\ref{eqn:3.5}). End of proof.
\par
Remarks :
The properties of the commutator which we use above are same as
those used by Feynman. We also use the Leibniz rule II (\ref{eqn:Leibniz2}).
\par
The dimension of the space-time is irrelevant to our proof.
This point forms a contrast to Feynman's case.
In his proof he used the complete antisymmetric tensor $ \epsilon_{ijk} $,
which depends on the dimension of space(-time).
The signature of the metric $ \eta_{\mu \nu} $ is also irrelevant.
It is sufficient that the metric is symmetric and regular.
\par
The parameter $ \tau $ is introduced as just a parameter. What could it be?
It may be identified with the proper-time but that is wrong.
First we do not use the condition
\begin{equation}
	\eta_{\mu \nu} \,
	\frac{\mbox{d} x^\mu}{\mbox{d} \tau} \,
	\frac{\mbox{d} x^\nu}{\mbox{d} \tau}
	= 1,
\label{eqn:proper-time}
\end{equation}
which claims $ \tau $ is the proper-time.
Secondly this condition is contradictory to the assumption (\ref{eqn:3.2}).
Eq. (\ref{eqn:proper-time}) implies
\begin{equation}
	[\: x^\lambda , \eta_{\mu \nu} \, \dot x^\mu \, \dot x^\nu \:] = 0.
\label{eqn:inc1}
\end{equation}
However Eq. (\ref{eqn:3.2}) implies
\begin{equation}
	[\: x^\lambda , \eta_{\mu \nu} \, \dot x^\mu \, \dot x^\nu \:] =
	- \frac{2 \, \k \hbar}{m} \, \dot x^\lambda .
\label{eqn:inc2}
\end{equation}
Apparently Eqs. (\ref{eqn:inc1}), (\ref{eqn:inc2}) are inconsistent.
\par
Our argument is not invariant under arbitrary reparametrization.
If we repara\-metrize $ \tau $ by another parameter $ \tau' $
as $ \tau = f(\tau') $, Eq. (\ref{eqn:3.2}) is transformed to
\begin{equation}
	m \, [ \: x^\mu, \dot x'^\nu \: ]
	= - \mbox{i} \hbar \, \dot f' \, \eta^{\mu \nu},
\label{eqn:3.23}
\end{equation}
where the dot and prime refer to differentiation with respect to $ \tau' $.
If the right-hand side of Eq. (\ref{eqn:3.23}) is not constant,
the derivation of Eq. (\ref{eqn:3.9}) cannot be justified.
Only permissible reparametrization is affine transformation,
$ \tau = a \tau' + b $.
In this case, metric, force and other quantities are transformed as follows :
\begin{eqnarray}
	&&
	\dot x'^\mu = a \, \dot x^\mu
	\nonumber\\
	&&
	\ddot x'^\mu = a^2 \, \ddot x^\mu
	\nonumber\\
	&&
	\eta'^{\mu \nu} = a \, \eta^{\mu \nu}
	\\
	&&
	F'^\mu = a^2 F^\mu
	\nonumber\\
	&&
	G'^\mu = a^2 G^\mu
	\nonumber\\
	&&
	F'^{\mu \nu} = a^2 F^{\mu \nu}.
	\nonumber
\end{eqnarray}
%
%
\section{General relativistic version}
\hspace{12pt}
Absence of gravity is one of the unsatisfactory points of Feynman's argument
and our previous one. However there is gravitation in our world!
How can we harmonize it with our framework?
\par
Here we propose one way to introduce it.
We take notice of the commutation relation (\ref{eqn:3.2}).
It seems natural to replace the Minkowskian metric $ \eta_{\mu \nu} $
by an arbitrary metric $ g_{\mu \nu}(x) $ to derive gravity.
We have found that this assumption leads to an anticipated result.
\\
Assumptions :
\begin{enumerate}
	\renewcommand{\labelenumi}{(\roman{enumi})}
	\item
		A particle is moving in $d$-dimensional
		space-time with coordinate $x^\mu(\tau)$
		$(\mu = 0, 1, \cdots, d-1)$, where $\tau$ is a parameter.
	\item
		Its coordinate and velocity $\dot x^\mu(\tau)$
		satisfy the commutation relations
		\begin{eqnarray}
			&&
			[ \: x^\mu, x^\nu \: ] = 0,
		\label{eqn:4.1}
			\\ &&
			m \, [ \: x^\mu, \dot x^\nu \: ]
			= - \mbox{i} \hbar \, g^{\mu \nu}(x),
		\label{eqn:4.2}
		\end{eqnarray}
		where the dot refers to the derivative with respect to $\tau$,
		and $ g^{\mu \nu}(x) $ is a metric of the space-time.
	\item
		It obeys the equation of motion
		\begin{equation}
			m \: \ddot x^\mu = F^\mu (x,\dot x).
		\label{eqn:4.3}
		\end{equation}
\end{enumerate}
Results :
\begin{enumerate}
	\renewcommand{\labelenumi}{(\roman{enumi})}
	\item
		The force $F^\mu(x,\dot x)$ can be written as
		\begin{equation}
			F^\mu(x,\dot x) =
			G^\mu(x) +
			\bra F^\mu_{\;\;\;\nu}(x) \, \dot x^\nu \ket -
			m \bra \Gamma^\mu_{\;\:\nu \rho} \,
			\dot x^\nu \, \dot x^\rho \ket .
		\label{eqn:4.4}
		\end{equation}
	\item
		The fields $G^\mu(x), F^\mu_{\;\;\;\nu}(x)$ satisfy
		\begin{eqnarray}
			&&
			\partial_\mu G_\nu - \partial_\nu G_\mu = 0,
		\label{eqn:4.5}
			\\ &&
			\partial_\mu F_{\nu \rho} +
			\partial_\nu F_{\rho \mu} +
			\partial_\rho F_{\mu \nu} =0.
		\label{eqn:4.6}
		\end{eqnarray}
	\item
		$\Gamma^\mu_{\;\:\nu \rho}(x)$ is a Levi-Civita connection,
		which is defined by
		\begin{equation}
			\Gamma_{\mu \nu \rho} = \frac12 (
			\partial_\rho g_{\mu \nu}+
			\partial_\nu  g_{\mu \rho}-
			\partial_\mu  g_{\nu \rho}).
		\label{eqn:4.7}
		\end{equation}
\end{enumerate}
Speaking after Dyson's fashion,
these results say that the only possible fields
that can consistently act on a quantum mechanical particle
are scalar, gauge and gravitational fields.
\par
Proof :
The tactics are almost the same as in the preceding section
but there is a problem concerning compatibility
between raising and lowering of tensor-indices
and ordering of operator-product.
The metric $ g_{\mu \nu}(x) $ is not a commutative operator.
Here we define lowering of the index of $ \dot x^\mu $ as
\begin{equation}
	\dot x_\mu = \bra g_{\mu \nu} (x) \, \dot x^\nu \ket ,
\label{eqn:4.8}
\end{equation}
using Weyl ordering $ \bra \cdots \ket $.
We write derivatives of a function $ f(x) $ as
\begin{eqnarray}
	&&
	\partial_\nu f = \frac{\partial f}{\partial x^\nu},
	\nonumber
	\\
	&&
\label{eqn:4.9}
	\\
	&&
	\partial^\nu f = g^{\nu \mu} \, \partial_\mu f.
	\nonumber
\end{eqnarray}
These definitions and Eqs. (\ref{eqn:4.1}),
(\ref{eqn:4.2}) give useful formulas
\begin{eqnarray}
	&&
	[\: \dot x_\nu, f(x) \:] =
	\frac{\k\hbar}m \: \partial_\nu f,
	\nonumber
	\\
	&&
\label{eqn:4.11}
	\\
	&&
	[\: \dot x^\nu, f(x) \:] =
	\frac{\k\hbar}m \: \partial^\nu f.
	\nonumber
\end{eqnarray}
For a general tensor field $ T^{\alpha \beta}(x) $,
we define lowering of its indices as usual,
\begin{equation}
	T_{\alpha \beta} = g_{\alpha \mu} \, g_{\beta \nu} \, T^{\mu \nu}.
\label{eqn:4.13}
\end{equation}
\par
Now we begin the proof.
Differentiating Eq. (\ref{eqn:4.2}) with respect to $\tau$
and using (\ref{eqn:4.3}), we obtain
\begin{equation}
	m \, [ \: \dot x^\mu, \dot x^\nu \: ]
	+ [ \: x^\mu , F^\nu \: ]
	= - \k\hbar \, \bra \partial_\rho g^{\mu \nu} \, \dot x^\rho \ket .
\label{eqn:4.14}
\end{equation}
We define $ W^{\mu \nu} $ as
\begin{equation}
	W^{\mu \nu}
	= - \frac{m^2}{\k\hbar} \, [ \: \dot x^\mu, \dot x^\nu \: ] .
\label{eqn:4.15}
\end{equation}
The Jacobi identity and Eqs. (\ref{eqn:4.2}), (\ref{eqn:4.11}) imply
\begin{eqnarray}
	[\: x^\lambda, W^{\mu \nu} \:]
	&=&
	- \frac{m^2}{\k\hbar} \,
	[\: x^\lambda, [\: \dot x^\mu, \dot x^\nu \:] \:]
	\nonumber
	\\
	&=&
	- \frac{m^2}{\k\hbar}
	\Bigl(
	  [\: [\: x^\lambda, \dot x^\mu \:], \dot x^\nu \:]
	+ [\: \dot x^\mu, [\: x^\lambda, \dot x^\nu \:] \:]
	\Bigr)
	\nonumber
	\\
	&=&
	m
	\Bigl(
	  [\: g^{\lambda \mu}, \dot x^\nu \:]
	+ [\: \dot x^\mu, g^{\lambda \nu} \:]
	\Bigr)
	\nonumber
	\\
	&=&
	- \k\hbar
	( \partial^\nu g^{\lambda \mu}
	- \partial^\mu g^{\lambda \nu} ).
\label{eqn:4.16}
\end{eqnarray}
Therefore if we put
\begin{equation}
	F^{\mu \nu} = W^{\mu \nu} - m \:
	\bra
	( \partial^\nu g^{\lambda \mu}
	- \partial^\mu g^{\lambda \nu} ) \dot x_\lambda
	\ket ,
\label{eqn:4.17}
\end{equation}
we obtain
\begin{equation}
	[\: x^\lambda, F^{\mu \nu} \:] = 0,
\label{eqn:4.18}
\end{equation}
which means that $ F^{\mu \nu} $ is a function of $ x $ only.
Eqs. (\ref{eqn:4.14}), (\ref{eqn:4.15}) and (\ref{eqn:4.17}) imply
\begin{eqnarray}
	[\: x^\mu, F^\nu \:]
	&=&
	\frac{\k\hbar}m \, F^{\mu \nu}
	+ \k\hbar \:
	\bra
	( \partial^\nu g^{\lambda \mu}
	- \partial^\mu g^{\lambda \nu}
	- \partial^\lambda g^{\mu \nu} ) \dot x_\lambda
	\ket
	\nonumber
	\\
	&=&
	\frac{\k\hbar}m \, F^{\mu \nu}
	+ 2 \, \k \hbar \:
	\bra
	\Gamma^{\nu \lambda \mu} \dot x_\lambda
	\ket ,
\label{eqn:4.19}
\end{eqnarray}
where we define $ \Gamma^{\nu \lambda \mu} $ as
\begin{equation}
	\Gamma^{\nu \lambda \mu}
	= - \frac12
	( \partial^\mu g^{\lambda \nu}
	+ \partial^\lambda g^{\mu \nu}
	- \partial^\nu g^{\lambda \mu} ),
\label{eqn:4.20}
\end{equation}
which is nothing but the Levi-Civita connection.
\par
Next we want to prove equation (\ref{eqn:4.6}).
For that purpose we should lower the indices of $ F^{\mu \nu} $.
It is easily seen that
\begin{eqnarray}
	- \frac{m^2}{\k\hbar} \,
	\bra \: [\: \dot x_\alpha, \dot x_\beta \:] \: \ket
	&=&
	- \frac{m^2}{\k\hbar} \,
	\bra \: [\:
	\bra g_{\alpha \mu} \dot x^\mu \ket,
	\bra g_{\beta \nu}  \dot x^\nu \ket
	\:] \: \ket
	\nonumber
	\\
	&=&
	- \frac{m^2}{\k\hbar} \,
	\bra
	g_{\alpha \mu} \, g_{\beta \nu} [\: \dot x^\mu, \dot x^\nu \:]
	+ g_{\alpha \mu} [\: \dot x^\mu, g_{\beta \nu} \:] \dot x^\nu
	\nonumber
	\\
	&&
	+ g_{\beta \nu} [\: g_{\alpha \mu}, \dot x^\nu \:] \dot x^\mu
	\ket
	\nonumber
	\\
	&=&
	\bra g_{\alpha \mu} \, g_{\beta \nu} W^{\mu \nu} \ket
	- m \,
	\bra
	\partial_\alpha g_{\beta \nu}   \, \dot x^\nu
	- \partial_\beta g_{\alpha \mu} \, \dot x^\mu
	\ket.
\label{eqn:4.21}
\end{eqnarray}
Eqs. (\ref{eqn:4.13}), (\ref{eqn:4.17}) and (\ref{eqn:4.21}) imply
\begin{eqnarray}
	F_{\alpha \beta}
	&=&
	g_{\alpha \mu} \, g_{\beta \nu} \, F^{\mu \nu}
	\nonumber
	\\
	&=&
	\bra g_{\alpha \mu} \, g_{\beta \nu} W^{\mu \nu} \ket
	- m \,
	\bra
	g_{\alpha \mu} \, g_{\beta \nu}
	( \partial^\nu g^{\lambda \mu}
	- \partial^\mu g^{\lambda \nu} ) \dot x_\lambda
	\ket
	\nonumber
	\\
	&=&
	- \frac{m^2}{\k\hbar} \,
	\bra \: [\: \dot x_\alpha, \dot x_\beta \:] \: \ket
	+ m \,
	\bra
	\partial_\alpha g_{\beta \nu}   \, \dot x^\nu
	- \partial_\beta g_{\alpha \mu} \, \dot x^\mu
	\ket
	\nonumber\\
	&&
	+ m \,
	\bra
	( g^{\lambda \mu} \, \partial_\beta  g_{\alpha \mu}
	- g^{\lambda \nu} \, \partial_\alpha g_{\beta  \nu} ) \dot x_\lambda
	\ket
	\nonumber
	\\
	&=&
	- \frac{m^2}{\k\hbar} \,
	\bra \: [\: \dot x_\alpha, \dot x_\beta \:] \: \ket,
\label{eqn:4.22}
\end{eqnarray}
when the second line turns to the third, the equation
$ 0 = \partial_\beta ( g_{\alpha \mu} \, g^{\lambda \mu} ) $
$   = ( \partial_\beta g_{\alpha \mu} ) g^{\lambda \mu} $
$   + g_{\alpha \mu} \, \partial_\beta g^{\lambda \mu} $
is used.
The Jacobi identity Eqs. (\ref{eqn:4.22}) and (\ref{eqn:4.11}) give
\begin{eqnarray}
	0
	&=&
	[\: \dot x_\mu , [\: \dot x_\nu , \dot x_\rho \:] \:] +
	[\: \dot x_\nu , [\: \dot x_\rho, \dot x_\mu  \:] \:] +
	[\: \dot x_\rho, [\: \dot x_\mu , \dot x_\nu  \:] \:]
	\nonumber\\
	&=&
	- \frac{\k\hbar}{m^2} \,
	\Bigl(
	[\: \dot x_\mu , F_{\nu  \rho} \:] +
	[\: \dot x_\nu , F_{\rho \mu } \:] +
	[\: \dot x_\rho, F_{\mu  \nu } \:]
	\Bigr)
	\nonumber\\
	&=&
	\frac{\hbar^2}{m^3} \,
	( \:
	\partial_\mu F_{\nu \rho} +
	\partial_\nu F_{\rho \mu} +
	\partial_\rho F_{\mu \nu}
	\: ),
\label{eqn:4.23}
\end{eqnarray}
which is just Eq. (\ref{eqn:4.6}).
\par
As in the previous proofs, we take Eq. (\ref{eqn:4.4})
as the definition of $ G^\mu $, that is
\begin{equation}
	G^\mu
	=
	F^\mu(x,\dot x)
	- \bra F^{\mu \nu}(x) \dot x_\nu \ket
	+ m \, \bra \Gamma^{\mu \nu \rho} \dot x_\nu \dot x_\rho \ket .
\label{eqn:4.24}
\end{equation}
Using Eqs. (\ref{eqn:4.24}), (\ref{eqn:4.19}) and (\ref{eqn:4.11})
in this order, we obtain
\begin{eqnarray}
	[\: x^\lambda, G^\mu \:]
	&=&
	[\: x^\lambda, F^\mu \:]
	- \bra F^{\mu \nu} [\: x^\lambda, \dot x_\nu \:] \, \ket
	\nonumber
	\\
	&&
	+ m \,
	\bra
	  \Gamma^{\mu \nu \rho} \, [\: x^\lambda, \dot x_\nu \:] \, \dot x_\rho
	+ \Gamma^{\mu \nu \rho} \, \dot x_\nu [\: x^\lambda, \dot x_\rho \:]
	\ket
	\nonumber
	\\
	&=&
	\frac{\k\hbar}m \, F^{\lambda \mu}
	+ 2 \, \k\hbar \,
	\bra \Gamma^{\mu \rho \lambda} \dot x_\rho \ket
	+ \frac{\k\hbar}m \, F^{\mu \nu} \, \delta^\lambda_{\;\:\nu}
	\nonumber
	\\
	&&
	- \k\hbar \,
	\bra
	  \Gamma^{\mu \nu \rho} \, \delta^\lambda_{\;\:\nu} \, \dot x_\rho
	+ \Gamma^{\mu \nu \rho} \, \dot x_\nu \, \delta^\lambda_{\;\:\rho}
	\ket
	\nonumber
	\\
	&=&
	0,
\label{eqn:4.25}
\end{eqnarray}
which says that $ G^\mu $ is a function of $ x $ only.
\par
The remaining task is to show equation (\ref{eqn:4.5}).
It takes tedious and long calculation,
so a reader who is not interested in the detail may skip
the following part of the proof.
Lowering of the index of Eq. (\ref{eqn:4.24}) gives
\begin{equation}
	G_\nu
	=
	\bra g_{\nu \alpha} \, F^\alpha \ket
	- \bra g_{\nu \alpha} \, F^{\alpha \beta} \, \dot x_\beta \ket
	+ m \bra g_{\nu \alpha} \, \Gamma^{\alpha \beta \gamma}
	\, \dot x_\beta \dot x_\gamma \ket.
\label{eqn:4.26}
\end{equation}
The first term of the right-hand side can be rewritten as
\begin{eqnarray}
	\bra g_{\nu \alpha} \, F^\alpha \ket
	&=&
	m \, \bra g_{\nu \alpha} \, \ddot x^\alpha \ket
	\nonumber
	\\
	&=&
	m \, \bra g_{\nu \alpha} \,
	\frac{\mbox{d}}{\mbox{d}\tau}
	\bra g^{\alpha \beta} \, \dot x_\beta \ket \ket
	\nonumber
	\\
	&=&
	m \,
	\bra
	g_{\nu \alpha} \, g^{\alpha \beta} \, \ddot x_\beta
	+
	g_{\nu \alpha} \, \partial^\gamma g^{\alpha \beta}
	\, \dot x_\beta \dot x_\gamma
	\ket
	\nonumber
	\\
	&=&
	m \,
	\bra
	\ddot x_\nu
	\ket
	+ m \,
	\bra
	g_{\nu \alpha} \, \partial^\gamma g^{\alpha \beta}
	\, \dot x_\beta \dot x_\gamma
	\ket.
\label{eqn:4.27}
\end{eqnarray}
In the second term of the last line, the indices $ \beta $ and $ \gamma $ of
$ \partial^\gamma g^{\alpha \beta} $ are symmetrized,
so it is rewritten using Eq. (\ref{eqn:4.20}) as
\begin{equation}
	\frac12 \, (
	\partial^\gamma g^{\alpha \beta} +
	\partial^\beta  g^{\alpha \gamma} )
	= - \Gamma^{\alpha \beta \gamma} +
	\frac12 \, \partial^\alpha g^{\beta \gamma}.
\label{eqn:4.28}
\end{equation}
Eqs. (\ref{eqn:4.26}), (\ref{eqn:4.27}) and (\ref{eqn:4.28}) imply
\begin{eqnarray}
	G_\nu
	&=&
	m \, \bra \ddot x_\nu \ket
	+
	\frac12 m \, \bra g_{\nu \alpha} \, \partial^\alpha g^{\beta \gamma}
	\, \dot x_\beta \dot x_\gamma \ket
	- \bra g_{\nu \alpha} \, F^{\alpha \beta} \, \dot x_\beta \ket
	\nonumber
	\\
	&=&
	m \, \bra \ddot x_\nu \ket
	+
	\frac12 m \, \bra \partial_\nu g^{\alpha \beta}
	\, \dot x_\alpha \dot x_\beta \ket
	- \bra F_{\nu \alpha} \, g^{\alpha \beta} \, \dot x_\beta \ket.
\label{eqn:4.29}
\end{eqnarray}
Using Eqs. (\ref{eqn:4.29}), (\ref{eqn:4.11}) and (\ref{eqn:4.22}), we obtain
\begin{eqnarray}
	&&
	[\: \dot x_\mu, G_\nu \:]
	\nonumber\\
	=
	&&
	m \bra \: [\: \dot x_\mu, \ddot x_\nu \:] \: \ket
	+
	\frac12 \k\hbar \bra \partial_\mu \partial_\nu g^{\alpha \beta}
	\, \dot x_\alpha \dot x_\beta \ket
	-
	\frac{\k\hbar}{2m} \bra \partial_\nu g^{\alpha \beta}
	( F_{\mu \alpha} \dot x_\beta + \dot x_\alpha F_{\mu \beta} ) \ket
	\nonumber\\
	&&
	- \frac{\k\hbar}m \bra \partial_\mu
	( F_{\nu \alpha} \, g^{\alpha \beta} ) \,
	\dot x_\beta \ket
	+
	\frac{\k\hbar}{m^2}
	\, F_{\nu \alpha} \, g^{\alpha \beta} \, F_{\mu \beta}
	\nonumber\\
	=
	&&
	m \bra \: [\: \dot x_\mu, \ddot x_\nu \:] \: \ket
	+
	\frac12 \k\hbar \bra \partial_\mu \partial_\nu g^{\alpha \beta}
	\, \dot x_\alpha \dot x_\beta \ket
	-
	\frac{\k\hbar}m \bra \partial_\nu g^{\alpha \beta} \cdot
	F_{\mu \alpha} \, \dot x_\beta \ket
	\nonumber\\
	&&
	- \frac{\k\hbar}m \bra \partial_\mu g^{\alpha \beta} \cdot
	F_{\nu \alpha} \, \dot x_\beta \ket
	-
	\frac{\k\hbar}m \bra \partial_\mu F_{\nu \alpha} \cdot
	g^{\alpha \beta} \, \dot x_\beta \ket
	+
	\frac{\k\hbar}{m^2}
	\, F_{\nu \alpha} \, g^{\alpha \beta} \, F_{\mu \beta} .
\label{eqn:4.30}
\end{eqnarray}
Finally
antisymmetrization with respect to the indices $ \mu $ and $ \nu $ gives
\begin{eqnarray}
	&&
	[\: \dot x_\mu, G_\nu \:] -
	[\: \dot x_\nu, G_\mu \:]
	\nonumber
	\\
	= &&
	m \bra \:
	[\: \dot x_\mu, \ddot x_\nu \:] -
	[\: \dot x_\nu, \ddot x_\mu \:]
	\: \ket
	- \frac{\k\hbar}m
	\bra \:
	( \partial_\mu F_{\nu \alpha}
	- \partial_\nu F_{\mu \alpha} ) g^{\alpha \beta} \, \dot x_\beta
	\: \ket
	\nonumber
	\\
	= &&
	m \bra \,
	\frac{\mbox{d}}{\mbox{d} \tau} [\: \dot x_\mu, \dot x_\nu \:]
	\: \ket
	- \frac{\k\hbar}m
	\bra \:
	( \partial_\mu F_{\nu \alpha}
	+ \partial_\nu F_{\alpha \mu} ) \dot x^\alpha
	\: \ket
	\nonumber
	\\
	= &&
	- \frac{\k\hbar}m \bra
	\frac{\mbox{d}}{\mbox{d} \tau} F_{\mu \nu}
	\ket
	- \frac{\k\hbar}m
	\bra \:
	( \partial_\mu F_{\nu \rho} + \partial_\nu F_{\rho \mu} ) \dot x^\rho
	\: \ket
	\nonumber
	\\
	= &&
	- \frac{\k\hbar}m
	\bra \:
	( \partial_\rho F_{\mu \nu}
	+ \partial_\mu  F_{\nu \rho}
	+ \partial_\nu  F_{\rho \mu} ) \dot x^\rho
	\: \ket .
\label{eqn:4.31}
\end{eqnarray}
Eqs. (\ref{eqn:4.6}), (\ref{eqn:4.11}) and (\ref{eqn:4.31})
imply (\ref{eqn:4.5}). End of proof.
\par
Remark :
Unfortunately we have not yet found the way to make
our formulation reparametrization-invariant.
%
%
\section{Non-Abelian gauge theory}
\hspace{12pt}
We can also bring the non-Abelian gauge field into our scheme.
However for this purpose we should admit more assumptions.
Our argument here is a special relativistic reconstruction of
C.R. Lee's work \cite{Lee}.
He generalized Feynman's proof to the case of non-Abelian gauge theory
in the nonrelativistic context.
\\
Assumptions :
\begin{enumerate}
	\renewcommand{\labelenumi}{(\roman{enumi})}
	\item
		A particle is moving in $d$-dimensional Minkowski
		space-time with coordinate $ x^\mu (\tau) $
		$ ( \mu = 0, 1, \cdots, d-1 ) $.
		And the particle carries isospin (or color) $ I^a (\tau) $
		$ ( a = 1, \cdots, n ) $.
		$ I^a $ 's are linearly independent operators.
	\item
		Its coordinate and velocity $\dot x^\mu(\tau)$ and isospin
		satisfy the commutation relations
		\begin{eqnarray}
			&&
			[ \: x^\mu, x^\nu \: ] = 0,
		\label{eqn:5.1}
			\\ &&
			m \, [ \: x^\mu, \dot x^\nu \: ]
			= - \mbox{i} \hbar \, \eta^{\mu \nu},
		\label{eqn:5.2}
			\\ &&
			[\: I^a, I^b \:] = \k\hbar  \, f_c^{\; ab} I^c,
		\label{eqn:5.3}
			\\ &&
			[\: x^\mu, I^a \:] = 0,
		\label{eqn:5.4}
		\end{eqnarray}
		where $ f_c^{\; ab} $ 's are the structure constants of
		the gauge group.
	\item
		Its coordinate obeys the equation of motion
		\begin{equation}
			m \: \ddot x^\mu = F^\mu (x,\dot x,I),
		\label{eqn:5.5}
		\end{equation}
		where the force $ F^\mu $ is linear with respect to $ I^a $,
		that is,
		\begin{equation}
			F^\mu (x,\dot x,I) =
			\bra F_a^\mu (x,\dot x) \, I^a \ket.
		\label{eqn:5.6}
		\end{equation}
		And its isospin obeys the equation
		\begin{equation}
			\dot I^a -
			f_c^{\;ab} \bra A_{b \mu}(x) \, \dot x^\mu I^c \ket
			= 0,
		\label{eqn:5.7}
		\end{equation}
		where $ A_{a \mu}(x) $ is a gauge field.
\end{enumerate}
Results :
\begin{enumerate}
	\renewcommand{\labelenumi}{(\roman{enumi})}
	\item
		The force $F^\mu(x,\dot x,I)$ can be written as
		\begin{equation}
			F^\mu(x,\dot x,I) =
			G_a^\mu(x) \, I^a +
			\bra F^\mu_{a \; \nu}(x) \, I^a \dot x^\nu \ket.
		\label{eqn:5.8}
		\end{equation}
	\item
		The fields $G_a^\mu(x), F^\mu_{a \; \nu}(x)$ satisfy
		\begin{eqnarray}
			&&
			( D_\mu G_\nu - D_\nu G_\mu )_a = 0,
		\label{eqn:5.9}
			\\ &&
			(
			D_\mu  F_{\nu \rho} +
			D_\nu  F_{\rho \mu} +
			D_\rho F_{\mu \nu} )_a = 0,
		\label{eqn:5.10}
		\end{eqnarray}
		where $ D $ denotes covariant derivative with the gauge field
		$ A_{a \mu} $, for instance,
		\begin{equation}
			( D_\mu F_{\nu \rho} )_a
			=
			\partial_\mu F_{a \nu \rho}
			- f_a^{\; bc} A_{b \mu} F_{c \, \nu \rho}.
		\label{eqn:5.11}
		\end{equation}
	\item
		The field $ F_{a \mu \nu} $ is related to the gauge potential
		$ A_{a \mu} $ by
		\begin{equation}
			f_c^{\; ab}
			\Bigl(
			F_{a \mu \nu} -
			( \partial_\mu A_{a \nu} - \partial_\nu A_{a \mu}
			- f_a^{\; de} A_{d \mu} A_{e \nu} )
			\Bigr)
			= 0 .
		\label{eqn:5.12}
		\end{equation}
\end{enumerate}
If we could remove $ f_c^{\; ab} $ 's from Eq. (\ref{eqn:5.12}),
we would identify $ F_{a \mu \nu} $ with the field strength
of the gauge potential $ A_{a \mu} $.
\par
The set of Eqs. (\ref{eqn:5.7}) and (\ref{eqn:5.8}) is
known as Wong's equation \cite{Wong},
which describes the motion of a particle carrying isospin.
In other words, it is non-Abelian extension of the Lorentz law.
Eq. (\ref{eqn:5.7}) just says that the isospin is parallel-transported
along the trajectory of the particle under the influence of the gauge field.
\par
Proof :
The tactics are almost the same as in the previous sections.
Differentiating Eq. (\ref{eqn:5.2}) with respect to $\tau$
and using (\ref{eqn:5.5}), we obtain
\begin{equation}
	m \, [ \: \dot x^\mu, \dot x^\nu \: ] + [ \: x^\mu , F^\nu \: ] = 0 .
\label{eqn:5.13}
\end{equation}
We define $ F^{\mu \nu} $ by
\begin{eqnarray}
	F^{\mu \nu}
	&=&
	- \frac{m^2}{\k\hbar} \, [ \: \dot x^\mu, \dot x^\nu \: ]
	\nonumber
	\\
	&=&
	\frac{m}{\k\hbar} \, [ \: x^\mu , F^\nu \: ]
	\nonumber
	\\
	&=&
	\frac{m}{\k\hbar} \, \bra \, [ \: x^\mu , F_a^\nu \: ] \, I^a \, \ket,
\label{eqn:5.14}
\end{eqnarray}
where we put
\begin{equation}
	F_a^{\mu \nu} =
	\frac{m}{\k\hbar} \, [ \: x^\mu , F_a^\nu(x,\dot x) \: ] ,
\label{eqn:5.15}
\end{equation}
therefore $ F_{\mu \nu} = \bra F_{a \mu \nu} \, I^a \ket $.
The Jacobi identity and Eq. (\ref{eqn:5.2}) imply
\begin{eqnarray}
	[ \: x^\lambda , F^{\mu \nu} \: ]
	&=&
	- \frac{m^2}{\k\hbar} \:
	[\: x^\lambda,[\:\dot x^\mu, \dot x^\nu \:]\:]
	\nonumber\\
	&=&
	- \frac{m^2}{\k\hbar} \:
	\Bigl(
		[\: [\: x^\lambda, \dot x^\mu \:], \dot x^\nu \:] +
		[\: \dot x^\mu, [\: x^\lambda , \dot x^\nu \:]\:]
	\Bigr)
	\nonumber\\
	&=&
	0,
\label{eqn:5.16}
\end{eqnarray}
which means that $ F^{\mu \nu} $ is independent of $ \dot x $
and $ F_a^{\mu \nu} $ is a function of $ x $ only.
\par
Differentiating Eq. (\ref{eqn:5.4}) with respect to $ \tau $ and
substituting (\ref{eqn:5.7}) and (\ref{eqn:5.2}), we obtain
\begin{eqnarray*}
	0
	&=&
	[\: \dot x^\mu, I^a \:] + [\: x^\mu, \dot I^a \:]
	\\
	&=&
	[\: \dot x^\mu, I^a \:] +
	f_c^{\; ab} [\: x^\mu, \bra A_{b \nu} \, \dot x^\nu I^c \ket \:]
	\\
	&=&
	[\: \dot x^\mu, I^a \:] -
	\frac{\k\hbar}m \,
	f_c^{\; ab} A_{b \nu} \, \eta^{\mu \nu} I^c ,
\end{eqnarray*}
which is rewritten as
\begin{equation}
	[\: \dot x_\mu, I^a \:]
	- \frac{\k\hbar}m \, f_c^{\; ab} A_{b \mu} I^c
	= 0
\label{eqn:5.17}
\end{equation}
by lowering the index $ \mu $.
{}From Eqs. (\ref{eqn:5.2}) and (\ref{eqn:5.17}), we derive a useful formula
\begin{eqnarray}
	[\: \dot x_\mu, \phi_a (x) I^a \:]
	&=&
	\frac{\k\hbar}m \,
	( \frac{\partial \phi_a}{\partial x^\mu}
	- f_a^{\; bc} A_{b \mu} \phi_c )
	\, I^a
	\nonumber
	\\
	&=&
	\frac{\k\hbar}m \,( D_\mu \phi )_a \, I^a
\label{eqn:5.18}
\end{eqnarray}
for functions $ \phi_a(x) \, ( a = 1,\cdots,n ) $.
\par
The Jacobi identity with (\ref{eqn:5.14}) and (\ref{eqn:5.18}) implies
\begin{eqnarray}
	0
	&=&
	[\: \dot x_\mu , [\: \dot x_\nu , \dot x_\rho \:] \:] +
	[\: \dot x_\nu , [\: \dot x_\rho, \dot x_\mu  \:] \:] +
	[\: \dot x_\rho, [\: \dot x_\mu , \dot x_\nu  \:] \:]
	\nonumber\\
	&=&
	- \frac{\k\hbar}{m^2} \,
	\Bigl(
	[\: \dot x_\mu , F_{\nu  \rho} \:] +
	[\: \dot x_\nu , F_{\rho \mu } \:] +
	[\: \dot x_\rho, F_{\mu  \nu } \:]
	\Bigr)
	\nonumber\\
	&=&
	\frac{\hbar^2}{m^3} \,
	( \,
	D_\mu F_{\nu \rho} +
	D_\nu F_{\rho \mu} +
	D_\rho F_{\mu \nu}
	\, )_a \, I^a,
\label{eqn:5.19}
\end{eqnarray}
and hence Eq. (\ref{eqn:5.10}) follows from the linear independence
of $ I^a $ 's.
\par
Next we show equation (\ref{eqn:5.12}). Using Eqs. (\ref{eqn:5.17}),
(\ref{eqn:5.14}), (\ref{eqn:5.2}) and (\ref{eqn:5.3}), we obtain
\begin{eqnarray}
	0
	&=&
	- \frac{m^2}{\k\hbar} \,
	\Bigl(
	[\: \dot x_\nu, [\: \dot x_\mu, I^a \:] \:]
	- \frac{\k\hbar}m \,
	f_c^{\; ab} [\: \dot x_\nu, A_{b \mu} I^c \:]
	\Bigr)
	\nonumber
	\\
	&&
	\nonumber
	\\
	&=&
	- \frac{m^2}{\k\hbar} \,
	\Bigl(
	[\: [\: \dot x_\nu, \dot x_\mu \:], I^a \:] +
	[\: \dot x_\mu, [\: \dot x_\nu, I^a \:] \:]
	\Bigr)
	\nonumber
	\\
	&&
	+ m \, f_c^{\; ab} \,
	\Bigl(
	[\: \dot x_\nu, A_{b \mu} \:] \, I^c +
	A_{b \mu} \, [\: \dot x_\nu, I^c \:]
	\Bigr)
	\nonumber
	\\
	&&
	\nonumber
	\\
	&=&
	[\: F_{\nu \mu}, I^a \:]
	- m \, [\: \dot x_\mu, f_c^{\; ab} A_{b \nu} I^c \:]
	\nonumber
	\\
	&&
	+ \k \hbar \, f_c^{\; ab} \,
	(
	\partial_\nu A_{b \mu} \, I^c
	+ A_{b \mu} \, f_e^{\; cd} A_{d \nu} I^e
	)
	\nonumber
	\\
	&&
	\nonumber
	\\
	&=&
	\k\hbar f_c^{\; ba} F_{b \nu \mu} I^c
	- \k\hbar f_c^{\; ab} \partial_\mu A_{b \nu} \, I^c
	- \k\hbar f_c^{\; ab} A_{b \nu} \, f_e^{\; cd} A_{d \mu} I^e
	\nonumber
	\\
	&&
	+ \k\hbar f_c^{\; ab} \,
	(
	\partial_\nu A_{b \mu} \, I^c
	+ f_e^{\; cd} A_{b \mu} \, A_{d \nu} I^e
	)
	\nonumber
	\\
	&&
	\nonumber
	\\
	&=&
	\k\hbar f_c^{\; ab} \, I^c
	(
	F_{b \mu \nu}
	- \partial_\mu A_{b \nu}
	+ \partial_\nu A_{b \mu}
	)
	\nonumber
	\\
	&&
	- \k\hbar f_c^{\; ab} \, f_e^{\; cd} \, I^e
	(
	A_{b \nu} \, A_{d \mu} -
	A_{b \mu} \, A_{d \nu}
	)
	\nonumber
	\\
	&&
	\nonumber
	\\
	&=&
	\k\hbar f_c^{\; ab} \, I^c
	(
	F_{b \mu \nu}
	- \partial_\mu A_{b \nu}
	+ \partial_\nu A_{b \mu}
	)
	\nonumber
	\\
	&&
	- \k\hbar \,
	( f_c^{\; ab} \, f_e^{\; cd} - f_c^{\; ad} \, f_e^{\; cb} ) \, I^e
	A_{b \nu} \, A_{d \mu} .
\label{eqn:5.20}
\end{eqnarray}
Concerning the last term, the Jacobi identity implies
\begin{eqnarray}
	- \hbar^2
	( f_c^{\; ab} \, f_e^{\; cd} - f_c^{\; ad} \, f_e^{\; cb} ) \, I^e \,
	&=&
	[\: [\: I^a, I^b \:], I^d \:] - [\: [\: I^a, I^d \:], I^b \:]
	\nonumber
	\\
	&=&
	[\: [\: I^a, I^b \:], I^d \:] + [\: I^b, [\: I^a, I^d \:] \:]
	\nonumber
	\\
	&=&
	[\: I^a, [\: I^b, I^d \:] \:]
	\nonumber
	\\
	&=&
	- \hbar^2
	f_c^{\; ae} \, f_e^{\; bd} \, I^c .
\label{eqn:5.21}
\end{eqnarray}
Therefore the last term of Eq. (\ref{eqn:5.20}) becomes
\begin{eqnarray}
	( f_c^{\; ab} \, f_e^{\; cd} - f_c^{\; ad} \, f_e^{\; cb} ) \, I^e \,
	A_{b \nu} \, A_{d \mu}
	&=&
	f_c^{\; ae} \, f_e^{\; bd} \, I^c \,
	A_{b \nu} \, A_{d \mu}
	\nonumber
	\\
	&=&
	- f_c^{\; ab} \, f_b^{\; de} \, I^c \,
	A_{e \nu} \, A_{d \mu}.
\label{eqn:5.22}
\end{eqnarray}
Equations (\ref{eqn:5.20}) and (\ref{eqn:5.22}) give (\ref{eqn:5.12}).
\par
As before, we take Eq. (\ref{eqn:5.8}) as the definition of $ G^\mu $, that is,
\begin{equation}
	G^\mu
	= G_a^\mu I^a
	= F^\mu (x,\dot x) - \bra F_a^{\mu \nu}(x) \, \dot x_\nu I^a \ket.
\label{eqn:5.23}
\end{equation}
Again, $ G_a^\mu $ might depend on $x$ and $\dot x$,
but using Eqs. (\ref{eqn:5.14}), (\ref{eqn:5.1}), (\ref{eqn:5.2})
and (\ref{eqn:5.4}), we get
\begin{eqnarray}
	[\: x^\lambda, G^\mu \:]
	&=&
	[\: x^\lambda, F^\mu \:] -
	\bra F^{\mu \nu} [\: x^\lambda, \dot x_\nu \:] \ket
	\nonumber
	\\
	&=&
	\frac{\k\hbar}m \, F^{\lambda \mu} +
	\frac{\k\hbar}m \, F^{\mu \nu} \, \delta^\lambda_{\;\;\nu}
	\nonumber
	\\
	&=&
	0,
\label{eqn:5.24}
\end{eqnarray}
which says that $ G_a^\mu $ is also a function of $x$ only.
\par
The remaining task is to show equation (\ref{eqn:5.9}).
Eqs. (\ref{eqn:5.18}) and (\ref{eqn:5.14}) imply
\begin{eqnarray*}
	&&
	[\: \dot x_\mu, G_\nu \:]
	\\
	= &&
	[\: \dot x_\mu, F_\nu \:] -
	\bra \: [\: \dot x_\mu, F_{\nu \rho} \:] \dot x^\rho \: \ket -
	\bra \: F_{\nu \rho} [\: \dot x_\mu, \dot x^{\rho} \:] \: \ket
	\\
	= &&
	m \, [\: \dot x_\mu, \ddot x_\nu \:] -
	\frac{\k\hbar}m \,
	\bra \: D_\mu F_{\nu \rho} \, \dot x^\rho \: \ket +
	\frac{\k\hbar}{m^2} \,
	\bra F_{\nu \rho} \, F_\mu^{\;\:\rho} \ket ,
\end{eqnarray*}
which leads to
\begin{eqnarray}
	&&
	[\: \dot x_\mu, G_\nu \:] -
	[\: \dot x_\nu, G_\mu \:]
	\nonumber
	\\
	= &&
	m [\: \dot x_\mu, \ddot x_\nu \:] -
	m [\: \dot x_\nu, \ddot x_\mu \:] -
	\frac{\k\hbar}m
	\bra \:
	( D_\mu F_{\nu \rho} - D_\nu F_{\mu \rho} ) \dot x^\rho
	\: \ket
	\nonumber
	\\
	&&
	+ \frac{\k\hbar}{m^2}
	\bra
	F_{\nu \rho} F_\mu^{\;\:\rho} - F_{\mu \rho} F_\nu^{\;\:\rho}
	\ket
	\nonumber
	\\
	= &&
	m \frac{\mbox{d}}{\mbox{d} \tau} [\: \dot x_\mu, \dot x_\nu \:] -
	\frac{\k\hbar}m
	\bra \:
	( D_\mu F_{\nu \rho} - D_\nu F_{\mu \rho} ) \, \dot x^\rho
	\: \ket
	\nonumber
	\\
	= &&
	- \frac{\k\hbar}m \frac{\mbox{d}}{\mbox{d} \tau} F_{\mu \nu}
	- \frac{\k\hbar}m
	\bra \:
	( D_\mu F_{\nu \rho} + D_\nu F_{\rho \mu} ) \, \dot x^\rho
	\: \ket
	\nonumber
	\\
	= &&
	- \frac{\k\hbar}m
	\bra \:
	( D_\rho F_{\mu \nu}
	+ D_\mu  F_{\nu \rho}
	+ D_\nu  F_{\rho \mu} ) \, \dot x^\rho
	\: \ket ,
\label{eqn:5.25}
\end{eqnarray}
with the aid of
\begin{eqnarray}
	\frac{\mbox{d}}{\mbox{d} \tau} ( F_{a \mu \nu} \, I^a )
	&=&
	\bra
	\partial_\rho F_{a \mu \nu} \, \dot x_\rho I^a
	+ F_{a \mu \nu} \, f_c^{\; ab} A_{b \rho} \dot x_\rho I^c
	\ket
	\nonumber
	\\
	&=&
	\bra (
	\partial_\rho F_{a \mu \nu}
	- f_a^{\; bc}  A_{b \rho} F_{c \mu \nu}
	) \, \dot x_\rho I^a
	\ket .
\label{eqn:5.26}
\end{eqnarray}
Eqs. (\ref{eqn:5.10}), (\ref{eqn:5.18}) and (\ref{eqn:5.25})
give (\ref{eqn:5.9}). End of proof.
\par
Remarks :
In our proof we should assume the existence of the gauge potential
$ A_{a \mu} $ in contrast with the Abelian case,
in which its existence is not an assumption but a result.
And we need the explicit form of the equation for the isospin,
that is Eq. (\ref{eqn:5.7}).
\par
We have not yet found the way to remove $ f_c^{\; ab} $ 's
from Eq. (\ref{eqn:5.12}).
\par
It may be possible to extend our argument to general relativistic formulation.
%
%
\section{Concluding remarks}
\hspace{12pt}
We have reviewed Feynman's proof and proposed
both a special relativistic and a general relativistic version
and extension to the non-Abelian gauge theory of his consideration.
However some questions remain.
\par
What on earth do these proofs show?
They may show to what degree commutation relations restrict allowable form
of the force.
However we have {\it not} used the noncommutativity of operators,
in other words,
we have {\it not} used the very definition $ [\, A,B \,] = AB - BA $.
We have used only the properties of $ [\; , \;] $, that is,
equations from (\ref{eqn:bilinearity}) to (\ref{eqn:Leibniz2}).
\par
If we persist in quantum mechanical context,
we should face the fact that
there is no satisfactory quantum theory of a single relativistic particle.
The commutation relation
\begin{equation}
	[\: x^\mu , m \dot x^\nu \:] = - \k\hbar \, \eta^{\mu \nu}
\end{equation}
implies that the eigenvalue of $ m \dot x^\mu $ runs over
the whole real numbers.
If we interpret $ m \dot x^\mu $ as energy-momentum $ p^\mu $ of the particle,
the problem of negative energy annoys us.
Even the on-shell condition $ p^\mu \, p_\mu = m^2 $ is not satisfied.
\par
If we take classical mechanical context,
we can make our consideration rather trivial. In the Lagrangian formalism,
the assumption (\ref{eqn:4.2}) may be replaced
by the statement that the Lagrangian is a quadratic function
with respect to the velocity, that is to say, the action is
\begin{equation}
	S = \int
	\Bigl(
	\frac12 m \, g_{\mu \nu}(x) \, \dot x^\mu \, \dot x^\nu
	+ A_\mu(x) \, \dot x^\mu
	+ \phi(x)
	\Bigr) \mbox{d} \tau.
\end{equation}
This assumption directly gives the results from
({\ref{eqn:4.4}) to ({\ref{eqn:4.7}).
But the above consideration is less attractive
because it demands much assumption, say,
existence of the potentials $ A_\mu, \phi $.
In our proof using the commutation relations,
their existence is not an assumption but a result.
\par
How seriously do we have to accept our result,
`` the only possible fields that can consistently act on
a quantum mechanical particle are scalar, gauge and gravitational fields '' ?
If we want other fields, there may be methods to introduce them.
A probable easy method is to replace the right-hand side of
Eq. (\ref{eqn:4.2}) by functions of both $ x $ and $ \dot x $.
In general, we can put
\begin{equation}
	m \, [ \: x^\mu, \dot x^\nu \: ]
	= - \mbox{i} \hbar \, \{
	  g^{\mu \nu}(x)
	+ h^{\mu \nu}_{\:\;\;\;\rho} (x)       \, \dot x^\rho
	+ k^{\mu \nu}_{\:\;\;\;\rho \sigma}(x) \, \dot x^\rho
	                                       \, \dot x^\sigma
	+ \cdots \} .
\end{equation}
We have not yet pursued extension by this method.
\par
A final question is this ; can we involve the spin in our argument?
The force which the electromagnetic and gravitational fields exert
over a particle with spin has different form from Eq. ({\ref{eqn:4.4}).
According to J.W. van Holten \cite{Holten1} \cite{Holten2}, it is
\begin{equation}
	m \, \ddot x^\mu
	= e g^{\mu \nu} F_{\nu \rho} \dot x^\rho
	- m \Gamma^\mu_{\;\: \nu \rho} \dot x^\nu \dot x^\rho
	+ \frac{e}{2m} g^{\mu \nu} ( \nabla_\nu F_{\rho \sigma} )
	S^{\rho \sigma}
	+ \frac 12 R^\mu_{\;\: \nu \rho \sigma} \dot x^\nu S^{\rho \sigma}
	,
\label{eqn:6.4}
\end{equation}
where $ e $ is electric charge of the particle,
which bas been absorbed in the definition of $ F_{\mu \nu } $
in the previous sections ;
$ \nabla $ denotes covariant derivative with the Levi-Civita connection
$ \Gamma^{\mu}_{\;\:\nu\rho} $ ;
$ R^\mu_{\;\: \nu \rho \sigma} $ is the Riemann curvature tensor ;
$ S^{\rho \sigma} $ is intrinsic angular momentum tensor, that is, spin.
We naively expect the Lie algebra of the (local) Lorentz group
\begin{equation}
	[\: S_{\mu \nu} , S_{\rho \sigma} \:]
	=
	- \k\hbar \, (
	  g_{\mu \rho} \, S_{\nu \sigma}
	- g_{\nu \rho} \, S_{\mu \sigma}
	- g_{\mu \sigma} \, S_{\nu \rho}
	+ g_{\nu \sigma} \, S_{\mu \rho}
	)
\end{equation}
to lead to Eq. ({\ref{eqn:6.4}).
Here we indicate the `local' Lorentz group parenthetically,
because the metric $ g_{\mu \nu}(x) $ depends on $ x $.
The derivation of Eq. (\ref{eqn:6.4}) is left for a future work.
%
%
\section*{Acknowledgments}
\hspace{12pt}
The author would like to thank Professor Y. Ohnuki,
 Professor S. Kitakado and Professor H. Ikemori for suggestive discussions.
He is deeply indebted to Prof. Ikemori and Dr. Tsujimaru
for their encouragement.
%
%

%
\end{document}